\documentstyle [12pt]{article}

\newcommand{\beq}{\begin{equation}}

\newcommand{\eeq}{\end{equation}}
\newcommand{\bea}{\begin{eqnarray}}
\newcommand{\eea}{\end{eqnarray}}
\newcommand{\nue}{\nu_e}
\newcommand{\nueb}{\bar{\nu}_e}
\newcommand{\numu}{\nu_\mu}

\newcommand{\nutau}{\nu_\tau}

\newcommand{\nus}{\nu_s}

\newcommand{\numt}{\nu_{\mu,\tau}}
\newcommand{\numtb}{\bar{\nu}_{\mu,\tau}}

\begin{document}
\baselineskip 7.4 mm

\def\thefootnote{\fnsymbol{footnote}}

\begin{flushright}
\begin{tabular}{l}
CERN-TH/97-5 \\
UPR-729-T \\
January, 1997 
\end{tabular}
\end{flushright}

\vspace{2mm}

\begin{center}

{\Large \bf 
Neutral current induced neutrino oscillations in a supernova 
}
\\ 
\vspace{8mm}

\setcounter{footnote}{0}

Alexander Kusenko\footnote{ email address:
kusenko@mail.cern.ch} \\
Theory Division, CERN, CH-1211 Geneva 23, Switzerland \\
and  \\
Gino Segr\`{e}\footnote{email address: segre@dept.physics.upenn.edu} 
\\
Department of Physics and Astronomy, University of Pennsylvania \\ 
Philadelphia, PA 19104-6396 \\

\vspace{12mm}

{\bf Abstract}
\end{center}

Neutral currents induced matter oscillations of electroweak-active 
(anti-)neutrinos to sterile neutrinos can explain the observed motion of
pulsars. In contrast to a recently proposed explanation of the pulsar
birth velocities based on the $\nu_{\mu,\tau} \leftrightarrow \nu_e$
oscillations, the heaviest neutrino (either active or sterile) 
would have to have mass of order several keV.

\vfill

\pagestyle{empty}

\pagebreak

\pagestyle{plain}
\pagenumbering{arabic}
\renewcommand{\thefootnote}{\arabic{footnote}}
\setcounter{footnote}{0}

\pagestyle{plain}

Neutrino physics continues to be at the forefront of  research in particle
physics and astrophysics.  There are many clues but the neutrino mass
matrix is still unknown and even the number of neutrino species is
still undetermined \cite{smirnov, ellis}.  Although the number of light 
electroweak active neutrinos is known from LEP, 
there are reasons to believe that massive sterile neutrinos may
exist and mix with the electroweak active neutrinos.  First, a
variety of experimental \cite{smirnov,ellis,pdg} and astrophysical \cite{ks} 
data can only be explained simultaneously if there are more than three neutrino
species.  Second, a standard model singlet fermion can naturally appear as
a modulino in models with broken supersymmetry.  

It was recently pointed out \cite{ks} that the proper motions of pulsars
can be explained if adiabatic neutrino oscillations take place inside a
cooling neutron star created in a supernova explosion.  The star's magnetic
field affects the location of the resonance in an up-down asymmetric manner
and causes the neutrinos of a
certain flavour to be emitted from different depths in different
directions.  Since the temperature inside a cooling neutron star depends
on the depth, the momentum distribution of the outgoing neutrinos will not
be spherically symmetric.  This can give the pulsar a sufficient recoil
velocity in good agreement with data \cite{ks}. 

If the pulsar motions are related to  neutrino oscillations,
they can be used as a  source of information about neutrino masses and
provide for  new astrophysical ``laboratory'' to complement the high-energy
experiments.  It is, therefore, important to examine variations of the
scenario proposed in Ref. \cite{ks} and their ramifications for neutrino
physics.  In this letter we concentrate on the effects  sterile neutrinos
can have on the motion of pulsars and estimate the magnitude of the pulsar 
birth velocity due to sterile-to-active neutrino oscillations.  The
main difference on the theoretical side between this effect and that
discussed in Ref. \cite{ks} is that neutral currents play a crucial role in
the oscillations of sterile neutrinos.

As neutrinos pass through matter, they experience an effective potential

\bea
V(\nus) & = & 0  \label{Vnus} \\
V(\nue)& = & -V(\nueb) =  V_0 \: (3 \, Y_e-1+4 \, Y_{\nue}) \label{Vnue} \\
V(\nu_{\mu,\tau}) & = & -V(\bar{\nu}_{\mu,\tau}) = V_0 \: ( Y_e-1+2 \, 
Y_{\nue}) \ 
+\ c_{_L}^{^Z} \: \frac{\vec{k} \cdot \vec{B}}{k} \label{Vnumu}
\eea
where $Y_e$ ($Y_{\nue}$) is the ratio of the number density of electrons
(neutrinos) to that of neutrons, $\vec{B}$ is the magnetic field, 
$\vec{k}$ is the neutrino momentum, $V_0=10 \: \rm{eV} \: (\rho/10^{14} g
\, cm^{-3} )$ and  

\beq
c_{_L}^{^Z}= \frac{e G_{_F}}{\sqrt{2}} \left ( \frac{3 N_e}{\pi^4} 
\right )^{1/3} 
\label{c}  
\eeq
The magnetic field dependent term in equation (\ref{Vnumu})
arises from a one-loop finite-density contribution \cite{ec} 
to the self-energy of a neutrino propagating in a magnetized 
medium\footnote{
We emphasize the difference between these chirality-preserving oscillations
and the spin and flavor precession of neutrinos in magnetic field studied,
{\it e.\,g.}, in Refs. \cite{fs,voloshin}, which can occur if the neutrino
magnetic moment is sufficiently large. Such oscillations can also have an
effect on the pulsar motions if the magnetic field is inhomogeneous 
\cite{voloshin}.}.   An excellent review of the neutrino ``refraction'' in
magnetized medium is found in Ref. \cite{smirnov}.

Only electrons contribute to the one-loop neutrino self-energy 
diagram with a charged current and an external photon source (magnetic
field). There are contributions from both electrons and protons to the  
the diagram with a neutral current and an external 
photon source.  In a neutral plasma in equilibrium, the neutral current
diagrams with electrons and protons cancel the charged current contribution 
\cite{ec}; therefore, there is no $(\vec{k} \cdot \vec{B})$ term in
equation (\ref{Vnue}).  In the case of $\numu$ and $\nutau$, the charged
current diagram is absent and the effective potential (\ref{Vnumu})
is magnetic field dependent.

The condition for resonant oscillation  $\nu_i \leftrightarrow \nu_j$ is 

\beq
\frac{m_i^2}{2 k} \: cos \, 2\theta_{ij} + V(\nu_i) = 
\frac{m_j^2}{2 k} \: cos \, 2\theta_{ij} + V(\nu_j)  
\label{resonance}
\eeq
where $\nu_{i,j}$ can be either a neutrino or an anti-neutrino. 

We consider a hierarchical mass matrix for neutrinos with 
$m(\nu_\tau)  \gg m(\nu_\mu) \gg m(\nu_e)$.  Most of the time after the
onset of the supernova explosion, the right-hand sides of equations
(\ref{Vnue}) and (\ref{Vnumu}) are negative because the deleptonization of 
nuclear matter during the first second causes the ratio $Y_e$ to drop from 
0.4 to about 0.1;  as the cooling of the neutron star proceeds, $Y_e$
further decreases to 0.04.  $Y_{\nu_e}\sim 0.07$ is also small. 

The sign of the mass difference determines
whether or not the resonant oscillations of a certain type can occur. 
If the sterile neutrino is heavier than other species, the
oscillations of the type  $\nu_s \leftrightarrow \bar{\nu_i}$ can take
place.  If, on the other hand,  $\nus$ is lighter than, {\it e.\,g.},
$\nutau$,  oscillations $\nu_s \leftrightarrow \nu_\tau$ are 
possible inside the neutron star.

The neutron star will receive a kick if the following two conditions
\cite{ks} are satisfied: (1)~the adiabatic oscillation $\nu_i
\leftrightarrow \nu_j$ occurs at a point inside the
$i$-neutrinosphere but outside the $j$-neutrinosphere; and (2)~the
difference $[V(\nu_i)-V(\nu_j)]$ contains a piece that depends on the
relative orientation of the magnetic field $\vec{B}$ and 
the momentum of the outgoing neutrinos, $\vec{k}$.  
If the first condition is satisfied,  the effective
neutrinosphere of $\nu_j$ will coincide with the surface formed by the
points of resonance.  The second condition ensures that this surface  
(a ``resonance-sphere'') will be deformed by the magnetic field in such a
way that it will be further from the center of the star when $(\vec{k}
\cdot \vec{B}) > 0$, and nearer when $(\vec{k} \cdot
\vec{B})<0$.  The average momentum carried away by the neutrinos depends on
the temperature of the region from which they exit.  The deeper inside
the star, the higher is the temperature.  Therefore, neutrinos coming
out in different directions will carry momenta which depend on the
relative orientation of  $\vec{k}$ and $\vec{B}$.  
This causes the asymmetry in  momentum distribution.  An $1\%$ asymmetry
is sufficient to generate  birth velocities of pulsars consistent with
observation \cite{ks}.  

Since the sterile neutrinos have a zero-radius neutrinosphere,  $\nu_s
\leftrightarrow \numtb$  oscillations can be the cause  of the pulsar 
motions if $m(\nus) > m(\numt)$.
If, on the other hand, $m(\nus) < m(\numt)$,  
$\nu_s \leftrightarrow \numt$ oscillations can play the same role. 
We emphasize that oscillations between a sterile neutrino and an electron
(anti-) neutrino are irrelevant for the recoil velocity of a pulsar.  We
will come back to this point below when we discuss the constraints from the
supernova 1987A.  

In the case of active neutrino oscillations, the magnitude of the ``kick''
was found \cite{ks} to be 

\begin{equation}
\frac{\Delta k}{k} = \frac{e}{3 \pi^2} \: \left ( \frac{\mu_e}{T}
\frac{dT}{dN_e} \right) B, 
\label{dk}
\end{equation}

The following changes occur for the active to sterile neutrino
oscillations: $N_e \equiv Y_e N_n$ is replaced by $N_n/2$, and 
there is an overall factor of 2 because, for a hierarchical mass matrix, 
the oscillations of both $\nu_\mu$ and $\nu_\tau$  occur at  nearby
points, both subject to the asymmetry in the magnetic field.  Therefore, 
for the neutral current induced oscillations, the size of the asymmetry in
momentum distribution is

\begin{equation}
\frac{\Delta k}{k} = \frac{4 e}{3 \pi^2} \: \left ( \frac{\mu_e}{T}
\frac{dT}{dN_n} \right) B, 
\label{dk1}
\end{equation}

To calculate the derivative in (\ref{dk1}), we use the relation between the
density and the temperature of a non-relativistic Fermi gas: 

\begin{equation}
N_n=\frac{2(m_n T)^{3/2}}{\sqrt{2} \pi^2}
\int \frac{\sqrt{z} dz}{e^{z-\mu_n/T}+1} 
\label{fermi}
\end{equation}
where $m_n$ and $\mu_n$ are the neutron mass and chemical potential. 
The derivative $(dT/dN_n)$ can be computed from (\ref{fermi}).  Finally,

\begin{equation}
\frac{\Delta k}{k} = \frac{4 e\sqrt{2}}{\pi^2} \: 
\frac{\mu_e \mu_n^{1/2}}{m_n^{3/2}T^2} \ B. 
\label{dk2}
\end{equation}

It is instructive to compare the magnitude of this asymmetry to that 
produced by the active neutrino oscillations \cite{ks}.  The latter agreed
well with the observed velocities of pulsars for $B\sim 10^{14}$ G.  
The right-hand side of equation (\ref{dk}) is different from equation
(\ref{dk2}) by the factor $X=4 \sqrt{2 \mu_n} \mu_e/m_n^{3/2}$, which is
maximized when the oscillations take place in the dense interior of the
star where the density is of order $10^{14} \, {\rm g\,cm^{-3}}$.  This 
corresponds to $N_n \sim (100\, {\rm MeV})^3 \approx (4/3 \sqrt{2} \pi^2)
(\mu_n m_n)^{3/2}$.  Therefore, $X \le 0.15$.  If the sterile neutrino
oscillatiosn are to explain the observed birth velocities of pulsars, the
magnetic field inside the star must be at least a factor of 6 greater than
that needed for the active neutrino oscillations to produce the required
asymmetry, or of order $B \sim 10^{15}$G. 

We conclude, therefore, that the active $\leftrightarrow$ sterile neutrino
oscillations, biased by the magnetic field, can result in the kick
velocities of order $\sim 500$ km/s if the resonant conversion takes place
at densities of order $\sim 10^{14}$ g/cm$^3$, one neutrino has a mass in
the keV range, and the magnetic field is $B\sim 10^{15} G$.  This requires
the mass  of one of the neutrinos to be in the $3 \ \rm{keV\ to} \ 10 \
\rm{keV}$ range.  This value of $B$ is, as was stated earlier, an order of
magnitude larger than the $B$ field needed in (\ref{dk}), but is still an
order of magnitude smaller than the $B$ field needed to begin to explain the
kick by other weak interactions \cite{voloshin,others}.  Of course, it is
not obvious that such large magnetic fields are possible inside a neutron
star. 

The mixing angle can be very small, because the adiabaticity
condition is satisfied if 

\begin{equation}
l_{osc} \approx \left (\frac{1}{2\pi} \ 
\frac{\Delta m^2}{2 k} \ sin \, 2 \theta
\right )^{-1} \approx \frac{10^{-2} \: {\rm cm}}{sin \, 2 \theta }
\end{equation}
is smaller than the typical scale of the density variations.  
Thus the oscillations remain adiabatic as long as $sin^2 \, 2 \theta > 
10^{-8}$.  

Since the sterile neutrino oscillations are assumed to occur inside the
$\bar{\nu}_e$-sphere, the flux of electron anti-neutrinos observed by 
IMB and Kamiokande \cite{SN1987A} is not affected as long as the sterile
neutrinos don't overcool the star.  
In the mass range of interest, $\Delta m \sim 3 - 10 \ \rm{keV}$, there is
an upper limit \cite{kimmo} on the mixing angle of the sterile neutrino and
the electron neutrino, $sin^2 \, 2 \theta_{e,s} > 10^{-6}$.  For larger
mixing angles, too big of a fraction of the energy is carried off by 
the sterile neutrinos, and the flux of electron neutrinos diminishes, in
contradiction with the observation of $\nueb$ events following SN1987A by
IMB and Kamiokande \cite{SN1987A}. 

Some comments are in order. The above estimate of the asymmetry in the
momentum of the outgoing neutrinos is reliable when only a small fraction
of the total energy is emitted as sterile neutrinos.  In the case
of active neutrino oscillations \cite{ks}, only $1/6$ of all neutrino types
exhibit the asymmetry, while the temperature distribution is determined by
the emission of all six.  The effect of the change in the position of the 
resonance on the overall temperature distribution is then small (next order
in $1/6$ treated as a small perturbation on the isotropic flux) and, to
first approximation, can be neglected.  In the case of active to sterile
neutrino oscillations, it is typically $1/3$ of the total neutrino flux
that is asymmetric.  This is because, for the hierarchical mass matrix with
$m(\nus) \gg m(\nu_{e,\mu,\tau})$, the $\nu_s \leftrightarrow
\bar{\nu}_\tau$ oscillations will take place at approximately the same 
density as the $\nu_s \leftrightarrow \bar{\nu}_\mu$ oscillations.
However, it is still reasonable to use the same approximation as before. 

Another possible caveat should be mentioned.  The temperature and density 
change rapidly in the vicinity of the neutrinospheres,  where the
cooling takes place.  Therefore, in calculating the momentum anisotropy in 
Ref. \cite{ks}, one could safely neglect the change in the luminous area 
when the position of the resonance changed slightly.  In general, the
total momentum is proportional to $T^4 \times (\sf{area})$, but as long as
the temperature varies faster than $1/R^2$, the area factor can be
considered constant.  Sterile neutrino oscillations may occur, however, 
deeper inside the neutron star, where the change in temperature is not so 
rapid, and the changes in the luminous area can modify the above estimate. 

We would like to thank K.~Kainulainen, P.~Langacker, and  A.~Smirnov 
for helpful discussions.  This work was supported in part by the
U.~S.~Department of Energy Contract No. DE-AC02-76-ERO-3071.


\begin{thebibliography}{99}

\bibitem{smirnov} A. Yu. Smirnov, Plenary talk given at 28th International
 Conference on High  energy physics, 25 - 31 July 1996, Warsaw, Poland
 (hep-ph/9611465).  

\bibitem{ellis} J. Ellis, Talk given at 17th International Conference
on Neutrino Physics and Astrophysics (Neutrino 96), Helsinki, Finland,
13-20 June 1996 (hep-ph/9612209).  

\bibitem{pdg} Particle Data Group, {\it Review of Particle Properties},
Phys. Rev. {\bf D54} (1996) 1. 

\bibitem{ks} A. Kusenko and G. Segr\`e, Phys. Rev. Lett. {\bf 77} 
(1996) 4872.  

\bibitem{ec} S.~Esposito and G.~Capone, Z.~Phys. {\bf C70} (1996) 55;  
J.~C.~D'Olivo, J.~F.~Nieves and P.~B.~Pal, Phys. Rev. {\bf D40} 
(1989) 3679; J.~C.~D'Olivo and J.~F.~Nieves, Phys. Lett. {\bf 
B383} (1996) 87.

\bibitem{fs} K.~Fujikawa and R.~Shrock, Phys. Rev. Lett. {\bf 45}  
(1980) 963; M.~B.~Voloshin, M.~I.~Vysotskii and L.~B.~Okun', 
Sov. Phys. JETP {\bf 64} (1987) 446 [Zh. Eksp. Teor. Fiz. {\bf 91} (1986)
754]; C.-S.~Lim and W.~J.~Marciano, Phys. Rev. {\bf D} (1988) 1368.

\bibitem{voloshin} M. B. Voloshin, Phys. Lett. {\bf B209} (1988) 360. 

\bibitem{others}  N.~N.~Chugai, Pis'ma Astron. Zh. {\bf 10}, 210 (1984); 
O.~F.~Dorofeev, V.~N.~Rodionov and I.~M.~Ternov,  
Sov. Astron. Lett. {\bf 11}, 123 (1985); A.~Vilenkin, Astrophys. J. 
{\bf 451}, 700 (1995);  C.~J.~Horowitz and J.~Piekarewicz, IU-NTC-96-18
(hep-ph/9701214). 

\bibitem{SN1987A} K. Hirata {\it et al.}, Phys. Rev. Lett. {\bf 58} (1987) 
1490; R.~M.~Bionta {\it et al.}, Phys. Rev. Lett. {\bf 58} (1987) 1494.

\bibitem{kimmo} K. Kainulainen, J.~Maalampi and J.~T.~Peltoniemi, 
Nucl. Phys. {\bf B358} (1991) 435;  E.~W.~Kolb, R.~N.~Mohapatra,
V.~L.~Teplitz, UMD-PP-96-93 (hep-ph/9605350).  

\end{thebibliography}
\end{document}